\documentclass[showpacs,aps,pra,superscriptaddress]{revtex4-1}

\usepackage{amsmath}
\usepackage{bm}
\usepackage{ucs}
\usepackage{geometry}
\usepackage{graphicx}
\usepackage{epstopdf} 
\usepackage{refcount}
\usepackage{array}

\graphicspath{{./Figures/}}
\begin{document}
\title{Snake instability of dark solitons across the BEC-BCS crossover: an effective field theory perspective}
\author{G. Lombardi}
\email{giovanni.lombardi@uantwerpen.be}
\affiliation{TQC, Universiteit Antwerpen, Universiteitsplein 1, B-2610 Antwerpen, Belgium}
\author{W. Van Alphen}
\email{wout.vanalphen@uantwerpen.be}
\affiliation{TQC, Universiteit Antwerpen, Universiteitsplein 1, B-2610 Antwerpen, Belgium}
\author{S. N. Klimin}
\affiliation{TQC, Universiteit Antwerpen, Universiteitsplein 1, B-2610 Antwerpen, Belgium}
\affiliation{Department of Theoretical Physics, State University of Moldova, 2009 Chi\unichar{537}in\unichar{259}u, Moldova}
\author{J. Tempere}
\affiliation{TQC, Universiteit Antwerpen, Universiteitsplein 1, B-2610 Antwerpen, Belgium}
\affiliation{Lyman Laboratory of Physics, Harvard University, Cambridge, MA 02138, USA}
\begin{abstract} 
In the present article the snake instability mechanism for dark solitons in superfluid Fermi gases is studied in the context of a recently developed effective field theory [Eur. Phys. J. B {\bf{88}}, 122 (2015)]. This theoretical treatment has proven to be suitable to study stable dark solitons in \textit{quasi-}1D setups across the BEC-BCS crossover. In this manuscript the nodal plane of the stable soliton solution is perturbed by adding a transverse modulation. The numerical solution of the system of coupled nonlinear differential equations describing the amplitude of the perturbation leads to the instability spectra which are calculated for a wide range of interaction regimes and compared to other theoretical predictions. The maximum transverse size that the atomic cloud can have in order to preserve the stability is estimated, and the effects of spin-imbalance on this critical length are examined, revealing a stabilization of the soliton with increasing imbalance.
\end{abstract} 
\maketitle
\section{Introduction}
Solitons are solitary waves that constitute the solution of a wide range of differential equations involving an interplay between nonlinearity and dispersive effects. They have been examined in many physical systems including optics, fluid mechanics, plasmas and ultracold gases.
In recent years, dark solitons were experimentally observed in both Bose Einstein condensates \cite{EXPDenschlag,EXPBurgerBongs,EXPAndersonHaljian,EXPBeckerStellmer} and Fermi superfluids, where the experimental realization proved to be less straightforward: initially \cite{EXPYefsah} soliton-like defects were observed, but their long-lifetime and high effective mass lead to the conclusion \cite{EXPKu} that what had been detected were in fact solitonic vortices, product of the decay of a planar soliton. Finally in \cite{EXPKu2} the entire evolution and decay of a dark soliton was experimentally observed. In ultracold quantum systems dark solitons manifest themselves as localized density dips propagating at constant velocity on a stable background. In correspondence to these defects the order parameter shows a dip in the amplitude profile and a jump in the phase profile. \\
From a theoretical point of view dark solitons were widely examined both in BECs \cite{THKonotopPitaevskii,THFrantzeskakis} and Fermi superfluids \cite{THAntezzaDalfovo,THSpuntarelliPieriStrinati,THLiaoBrand,THScottDalfovo}. For what concerns the stability of the soliton, it was demonstrated \cite{THMuryshev,THFeder,THBrandReinhardt} that, while a soliton in 1D configurations is stable, the presence of a transverse dimension provokes its decay through the snake instability mechanism. The name snake instability \cite{THZakharovRubenchik} comes from the consideration that in the decay process the depletion plane starts to oscillate until the defect loses its soliton character: the cascade of solitonic excitations arising from the snaking of a planar soliton was observed experimentally both in BECs \cite{EXPDonadello} and in Fermi superfluids \cite{EXPKu2}. Theorists have analyzed the snake instability mechanism in fermionic systems by employing different methods, e.g. hydrodynamic approximation, RPA approach, the numerical solution of the time-dependent Bogoliubov-de Gennes equations \cite{THCetoliBrand}, and calculations \cite{THMunozBrand} based on a coarse-grained version of the BdG equations introduced in \cite{THSimonucciStrinati}.\\ In the present paper we study 
the snake instability by using a recently developed effective field theory \cite{THKTLDEpjB} capable of describing Fermi superfluids across the BEC-BCS crossover regime in a wide temperature domain. This EFT has been already employed in the description of the stable soliton solution in (quasi-)1D setups in different regimes of temperature and imbalance \cite{THKTDPrA,THLvAKTPrA}. One goal of our work is to analyze the spectrum of the instability, and to compare the results of the EFT with those of other theoretical approaches across the BEC-BCS crossover regime.
The snake instability is a long-wavelength phenomenon: the inverse of the maximum wavenumber $k_c$ for which the soliton is unstable, i.e. $k_c^{-1}$, can therefore give an estimate of the minimum transverse length that the atomic cloud must have in order for the decay to occur in experiment. Since in experiments with Fermi superfluids no stable solitons have been observed yet, the estimate of this quantity can help design future experimental observations. The study carried out in this work is based on the perturbation of stationary soliton solutions obtained from the analytic expressions for the phase and amplitude profiles of the order parameter derived in \cite{THKTDPrA,THLvAKTPrA} which are valid for a uniform system. The recent realization of box-like optical traps \cite{EXPGauntSchmidutz} provides the opportunity to test the predictions of the present work in experiment.\\
The high experimental control obtained in ultracold fermionic systems enables to tune a broad variety of parameters, from temperature to interaction strength. In addition, in 2006 physicists managed to experimentally engineer spin-imbalanced systems \cite{EXPZwierlein,EXPPartridge}, in which the populations of particles with spin-up and spin-down are uneven. From a theoretical point of view the difference in the spin-populations has been demonstrated to strongly affect the pairing mechanism \cite{THTKDPrA79}. In this work the effects of imbalance are analyzed with respect to the minimal transverse length $k_c^{-1}$ and the stability of the stationary soliton solution.


\section{Perturbative treatment} \label{pertreat}
The system under consideration is an ultracold Fermi gas in which particles in states with opposite pseudo-spin interact via an $s-$wave contact potential. In the context of a recently developed effective field theory \cite{THKTLDEpjB}, this system can be described across the BEC-BCS crossover regime in terms of the superfluid order parameter $\Psi$. The Euclidean-time action functional is given, in the natural units of  $\hbar=1$, $2m=1$, $E_F=1$, by
\begin{equation}
S(\beta)=\int_0^\beta\mathrm{d}\tau\int\mathrm{d}\bm{r}\left[\frac{D}{2}\left(\bar{\Psi}\frac{\partial \Psi}{\partial\tau}-\frac{\partial\bar{\Psi}}{\partial \tau}\Psi\right)+\mathcal{H}\right] \,,\label{Sbeta}
\end{equation}
where $\beta$ is the inverse temperature, and the Hamiltonian $\mathcal{H}$ is given by
\begin{equation}
\mathcal{H}=\Omega_s(\Psi)+\frac{C}{2m}\left|\nabla_{\bm{r}}\Psi\right|^2 -\frac{E}{2m}\left(\nabla_{\bm{r}}\left|\Psi\right|^2\right)^2 \,. \label{H}
\end{equation}
The analytic expressions for the coefficients $C, D, E$ of the EFT, and for the thermodynamic potential $\Omega_s$ are given in \citep{THKTLDEpjB,THLvAKTPrA} in terms of the order parameter $\Psi$, chemical potential $\mu$, imbalance parameter $\zeta$, and interaction parameter $(k_Fa_s)^{-1}$. To make this paper self-contained, the definitions of the EFT coefficients and a brief discussion about their relation to the system parameters are included in the appendix.
The regularized real-time Lagrangian density is 
\begin{equation}
\mathcal{L}=\mathrm{i}\frac{D}{2}\left(\bar{\Psi}\frac{\partial \Psi}{\partial t}-\frac{\partial\bar{\Psi}}{\partial t}\Psi\right)-\left(\mathcal{H}-\Omega_s(\Psi_\infty)\right)\,.
\end{equation}
where $\Psi_\infty$ is the value of the order parameter for a uniform system which can be obtained, given the values for temperature, interaction strength and imbalance, by solving the gap equation $\partial \Omega_s (\Psi)/\partial{\Psi}=0$. The subtraction of the term $\Omega_s(\Psi_\infty)$ means that in the present treatment the energy is always intended as the energy difference with respect to the value of the thermodynamic potential for the uniform system.\\ 
From the effective field Lagrangian, the equation of motion for the pair field $\Psi$ of the Fermi superfluid can be obtained, reading
\begin{equation}
i\tilde{D}(\vert\Psi\vert^2)\frac{\partial \Psi }{\partial t}=-\frac{C}{2m} \, \nabla _{\boldsymbol{r}}^{2}\Psi +\left( A(\vert\Psi\vert^2)+\frac{E}{m} \, \nabla _{\boldsymbol{r}}^{2}|\Psi |^{2}\right) \Psi  \,. \label{eqofmot}
\end{equation}
The coefficients $A$ and $\tilde{D}$ introduced in the last expression are defined in \eqref{A&D}.
In the one-dimensional (1D) case, equation \eqref{eqofmot} can be solved analytically and an exact solution $\Psi _{s}(x-v_{s}t)$ for a stable soliton that propagates with velocity $v_s$ can be found. The properties of such solitonic solutions have been thoroughly examined in \cite{THKTDPrA,THLvAKTPrA}.
This 1D treatment, however, does not capture the physics of the snaking mechanism because, in order for the instability to develop, a transverse direction is needed. To describe the deformation of the soliton plane that leads to its decay, a transverse perturbation is added to the stationary 1D soliton in the following way \cite{THKuznetsovTuritsyn}
\begin{equation}
\Psi (x,z,t)=\Psi _{s}(x-v_{s}t)+\Phi (x-v_{s}t,z,t)\, ,  \label{Psi}
\end{equation}%
where the perturbation $\Phi (x-v_s t,z,t)$ is assumed to be small. The space- and time-dependence of the correction is assumed to have the form $x-v_{s}t$, meaning that it propagates in the $x$ direction with velocity $v_s$ in the same way as the soliton does. The perturbation is further assumed to consist of a combination of plane wave components propagating in opposite directions:
\begin{equation}
\Phi (x-v_{s}t,z,t)=\phi _{1}(x-v_{s}t)e^{i(kz-\Omega t)}+\phi_{2}^*(x-v_{s}t)e^{-i(kz-\Omega^* t)}\label{Phi}
\end{equation}%
The next step is to insert this perturbed solution into the equation of motion \eqref{eqofmot} and to perform an expansion around the stationary solution up to first order in $\Phi$. From previous considerations \cite{THKTDPrA}, we know that the coefficients $C$ and $E$ can be kept constant and equal to their value in the uniform system case. On the other hand, the dependence of both $\tilde{D}$ and $\mathcal{A}$ on the order parameter has to be fully considered. A Taylor expansion of these two coefficients up to first order around the stationary solution leads to
\begin{equation}
\tilde{D}(|\Psi |^{2})=\tilde{D}(|\Psi _{s}|^{2})+\frac{\partial \tilde{D}(|\Psi _{s}|^{2})}{\partial |\Psi _{s}|^{2}}\left[ \left( \Psi _{s}^{\ast}\phi _{1}+\Psi _{s}\phi _{2}^{\ast }\right) e^{i(kz-\Omega t)}+\left( \Psi_{s}\phi _{1}^{\ast }+\Psi _{s}^{\ast }\phi _{2}\right) e^{-i(kz-\Omega t)}\right] +\ldots \label{Aexp}
\end{equation}%
\begin{equation}
A(|\Psi |^{2})=A(|\Psi _{s}|^{2})+\frac{\partial A(|\Psi _{s}|^{2})}{\partial |\Psi _{s}|^{2}}\left[ \left( \Psi _{s}^{\ast }\phi _{1}+\Psi_{s}\phi _{2}^{\ast}\right)e^{i(kz-\Omega t)}+\left( \Psi _{s}\phi_{1}^{\ast}+\Psi _{s}^{\ast }\phi _{2}\right) e^{-i(kz-\Omega t)}\right] +\ldots \label{Dexp}
\end{equation}
After inserting \eqref{Aexp} and \eqref{Dexp} into the equation of motion and expanding the temporal and spatial derivatives,  the terms of order zero in the perturbation can be collected, leading to
\begin{equation}
i\tilde{D}(\vert\Psi_s\vert^2)\frac{\partial \Psi_s }{\partial t}=-\frac{C}{2m} \, \nabla _{\boldsymbol{r}}^{2}\Psi_s +\left( \mathcal{A}(\vert\Psi_s\vert^2)+\frac{E}{m} \, \nabla _{\boldsymbol{r}}^{2}|\Psi_s |^{2}\right) \Psi_s 
\end{equation}%
which is, as expected, just the equation of motion for the stationary solution. From the selection of the terms that are linear in the perturbation, two coupled differential equations are obtained for the perturbation amplitudes $\phi_1$ and $\phi_2$:
\begin{align}
\alpha_1 \frac{\partial ^{2}\phi _{1}}{\partial x^{2}} - \alpha_2 \frac{\partial \phi _{1}}{\partial x}  +\alpha_3 (\Omega) \phi _{1} - \alpha_4 \frac{\partial ^{2}\phi _{2}}{\partial x^{2}} - \alpha_5 \frac{\partial \phi _{2}}{\partial x}  - \alpha_6 \phi _{2}=0 \label{eq:diff1}
\end{align}
\begin{align}
\alpha_1 \frac{\partial ^{2}\phi _{2}}{\partial x^{2}} - \alpha_2^* \frac{\partial \phi _{2}}{\partial x}  +\alpha_3^*(-\Omega) \phi _{2} - \alpha_4^* \frac{\partial ^{2}\phi _{1}}{\partial x^{2}} - \alpha_5^* \frac{\partial \phi _{1}}{\partial x}  - \alpha_6^* \phi _{1}=0 \label{eq:diff2}
\end{align}
where the coefficients $\alpha_j$, $j=1,2,3,4,5,6$ are defined as

\vspace*{0.1cm}
\begin{tabular}{>{\centering\arraybackslash}p{0.5\linewidth} @{\bigg|} >{\centering\arraybackslash}p{0.5\linewidth} }
$\displaystyle \alpha_1 = \frac{C}{2m}-\frac{E}{m}\vert\Psi_s\vert^2 $
&
$\displaystyle \alpha_2 = iv_{s}\tilde{D}_s + 2 \frac{E}{m} \Psi_s \frac{\partial \Psi_s^*}{\partial x}$ \\[0.5cm]
\multicolumn{2}{c}{$\displaystyle \alpha_3 = \Omega \tilde{D}_s -\frac{C}{2m}k^{2}-\partial_s \left(\vert\Psi_s\vert^2 \mathcal{A}_s \right)-iv_{s} \partial_s \tilde{D}_s \frac{\partial \Psi _{s}}{\partial x}\Psi _{s}^{\ast }  -\frac{E}{m} \frac{\partial^2 \vert\Psi_s\vert^2}{\partial x^2} - \frac{E}{m} \Psi_s \frac{\partial^2 \Psi_s^*}{\partial x^2} + \frac{E}{m} \vert\Psi_s\vert^2 k^2$} \\[0.5cm]
$\displaystyle \alpha_4 =  \frac{E}{m} \Psi_s^2 \frac{\partial^2 \phi_2}{\partial x^2} $
&
$\displaystyle \alpha_5 = 2 \frac{E}{m} \Psi_s \frac{\partial \Psi_s}{\partial x} \frac{\partial \phi_2}{\partial x} $\\[0.5cm]
\multicolumn{2}{c}{$\displaystyle \alpha_6 = \partial_s \mathcal{A}_s \Psi_s^2 +iv_{s} \partial_s\tilde{D}_s\frac{\partial \Psi _{s}}{\partial x}\Psi _{s} + \frac{E}{m}\Psi_s \frac{\partial^2 \Psi_s}{\partial x^2} - \frac{E}{m} \Psi_s^2 k^2$} \\[0.4cm]
\end{tabular}
In the last set of expressions we introduced the notations
\begin{align*}
F(\vert \Psi_s \vert^2)= F_s \, ,\qquad \frac{\partial F}{\partial \vert \Psi_s \vert^2} = \partial_s F
\end{align*}
(where $F$ can stand for $A$ or $\tilde{D}$).
\section{Results} \label{res}
\begin{figure}[h]\centering
\includegraphics[scale=0.6]{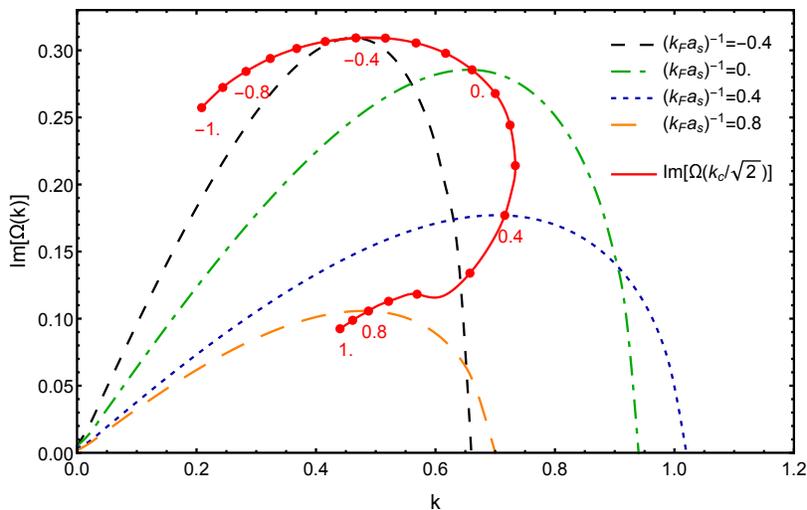}\caption{Dispersion relations for the snake instability in different interaction conditions across the BEC-BCS crossover, i.e. on the BCS side of the resonance at $(k_Fa_s)^{-1}=-0.4$ (black dashed line), at unitarity $(k_Fa_s)^{-1}=0$ (green dot-dashed line), in the near-BCS regime $(k_Fa_s)^{-1}=0.4$ (blue dotted line) and further towards the BEC limit at $(k_Fa_s)^{-1}=0.8$ (orange wide-dashed line). The full red line connects the values of $\text{Im}[\Omega(k)]$ calculated in $k=k_c/\sqrt{2}$ for different values of $(k_Fa_s)^{-1}$. The markers correspond to values of $(k_Fa_s)^{-1}$ ranging from $-1$ to $1$ in steps of $0.1$.}\label{fig:disp}
\end{figure}

\begin{figure}[h]\centering
\includegraphics[scale=0.6]{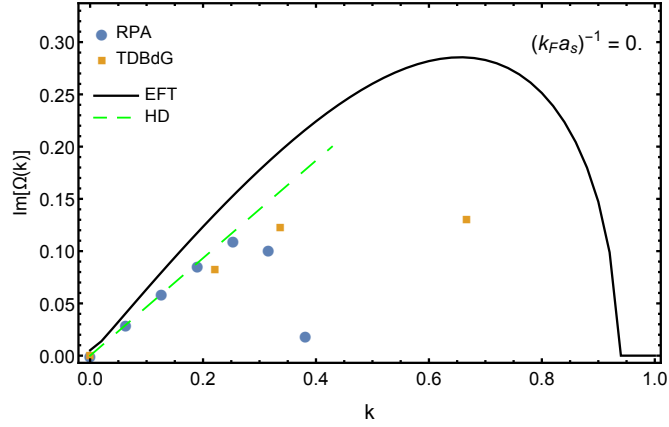}\caption{Dispersion relations for the snake instability at unitarity for $(k_Fa_s)^{-1}=0$. The full black line represents the EFT prediction and it is compared to the results of the hydrodynamic approximation (green dashed line), of the RPA (blue circles), and of the TDBdG simulations (orange squares) \cite{THCetoliBrand} }\label{fig:ka0}
\end{figure}

\begin{figure}[h]\centering
\includegraphics[scale=0.6]{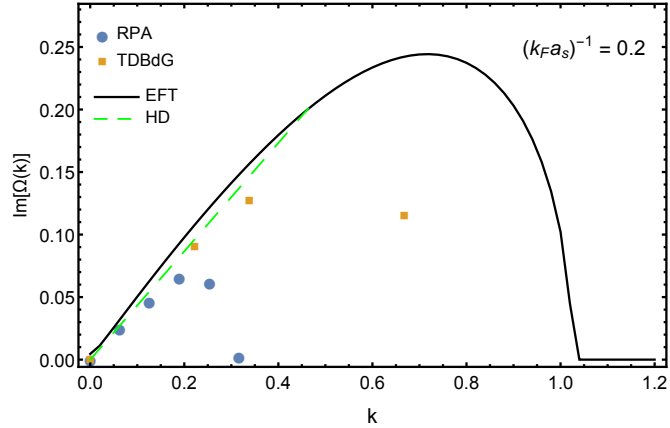}\caption{Dispersion relations for the snake instability in the near BEC regime for $(k_Fa_s)^{-1}=0.2$. The full black line represents the EFT prediction and it is compared to the results of the hydrodynamic approximation (green dashed line), of the RPA (blue circles), and of the TDBdG simulations (orange squares) \cite{THCetoliBrand} }\label{fig:ka02}
\end{figure}

\begin{figure}[h]\centering
\includegraphics[scale=0.6]{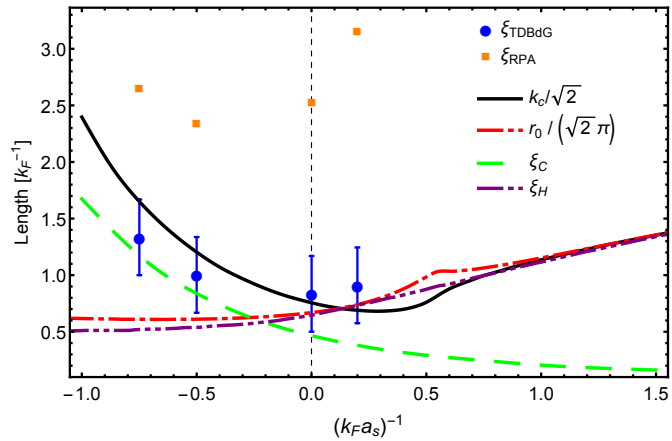}\caption{The EFT prediction for the minimum transverse dimension necessary for observing soliton decay through the snake instability (full black line) is compared to the results of the RPA (orange squares), of the TDBdG simulations (blue circles) \cite{THCetoliBrand}, and of the calculations by Mu\~noz Mateo and Brand \cite{THMunozBrand} (red dot-dashed line) based on the coarse-grained BdG theory \cite{THSimonucciStrinati}. In addition the BCS coherence length (green dashed line) and  BEC healing length (purple dot-dot-dashed line) are shown. The numerical factors are introduced to overcome differences in the definitions of the healing lengths between Refs. \cite{THMunozBrand,THCetoliBrand,THMuryshev} as discussed in \cite{Note1}.}\label{fig:Lcomp}
\end{figure}

\begin{figure}[h]\centering
\includegraphics[scale=0.6]{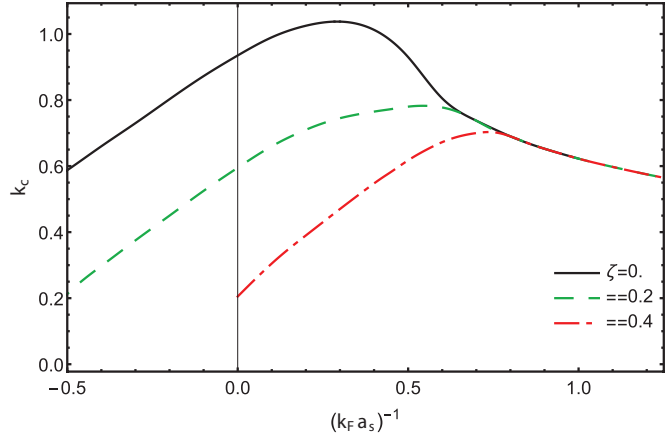}\caption{Critical wavenumber $k_c$ for the occurrence of the snake instability across the BEC-BCS crossover for different values of the imbalance parameter $\zeta$, i.e.\ $\zeta=0$ (full black line), $\zeta = 0.2$ (green dashed line) and $\zeta = 0.4$ (red dot-dashed line). The lines for $\zeta\neq 0$ do not cover the entire interaction domain due to the fact that in the presence of imbalance the superfluid state does not exist across the whole BEC-BCS crossover.}\label{fig:kcZ}
\end{figure}

From the system of coupled differential equations \eqref{eq:diff1} and \eqref{eq:diff2} one can obtain information about the perturbation's frequency spectrum $\Omega(k)$. In particular, the soliton solution will be unstable for every wavevector $k$ that corresponds to an imaginary value of the frequency. Therefore, the first goal of the present work is to analyze the imaginary part of the spectrum $\Omega(k)$ and obtain a description of the growth rate of the instability in different interaction regimes across the BEC-BCS crossover. To do this, the system of equations is approached as an eigenvalue problem of the form
\begin{equation}
\left( \begin{matrix}
W_{11} & W_{12} \\
W_{21} & W_{22}
\end{matrix} \right) %
\left( \begin{matrix}
 \phi_1 \\
 \phi_2
\end{matrix} \right) %
= %
\Omega \left( \begin{matrix}
 \phi_1 \\
 \phi_2
\end{matrix} \right)
\end{equation} 
and is solved numerically for the case of a stationary soliton ($v_S=0$) by approximating the derivatives with finite differences on a large space grid \cite{THColeMusslimani}. 
\\Figure \ref{fig:disp} shows the results for the imaginary part of the eigenvalues $\Omega(k)$ at $T=0.01T_F$ and for different values of the interaction parameter $(k_F a_s)^{-1}$. It is clear that the snake instability is a long-wavelength phenomenon that only exists up to a maximum wavenumber $k_c$ since the imaginary part of the frequency $\Omega$ is zero for $k>k_c$. The full red line interpolates between the values of $\text{Im}\left[\Omega(k)\right]$ calculated in $k=k_c/\sqrt{2}$. As predicted by Muryshev \textit{et al.} in the case of Bose Einstein condensates \cite{THMuryshev}, this line nicely connects the maxima of the dispersion relations for different $(k_Fa_s)^{-1}$. \\  
Figure \ref{fig:ka0} and \ref{fig:ka02} compare the results for $(k_F a_s)^{-1} = 0$ and $(k_F a_s)^{-1} = 0.2$ with the corresponding spectra that were calculated in Ref. \cite{THCetoliBrand}. There, the authors made use of three different approaches to analyze the spectra of the snake instability: a hydrodynamic approximation, the random-phase approximation (RPA) and the solution of the time-dependent Bogoliubov-de Gennes (TDBdG) equations. For what concerns the width of the band of unstable wavelengths, the latter method shows the best agreement with the EFT results. The RPA results on the other hand show a sharp decrease of $\text{Im}[\Omega]$, which might be caused by the necessary use of an energy cutoff in this type of calculations, an issue that does not occur in the presently used EFT. Another consequence of this cutoff is that the RPA method fails to find any imaginary frequency at all for $(k_F a_s)^{-1} > 0.2$. The hydrodynamic approximation, that describes a linear relation between $\Omega$ and $k$, is only expected to hold near $k=0$, where it indeed agrees quite well with the initial slope of the present results.\\
The existence of a minimum wavenumber $k_c$ for which $\text{Im}(\Omega)$ becomes zero implies that there exists a minimal transverse length the ultracold gas must have in order for the soliton to decay. If the transverse width is smaller than this minimal value, the soliton is expected to be stable. A good estimate for this critical length is given by the inverse of $k_c$. 
In figure \ref{fig:Lcomp} this quantity is compared to the RPA and TDBdG results of \cite{THCetoliBrand} as well as to the data from \cite{THMunozBrand} relative to a treatment based on the coarse-grained BdG equation introduced by Simonucci and Strinati \cite{THSimonucciStrinati}. Numerical factors have been introduced after a cross-comparison between Refs. \cite{THCetoliBrand,THMunozBrand,THMuryshev} in order to overcome the difference in the definitions of the healing lengths \footnote{In \cite{THMunozBrand} the quantity $r_0$ is defined as $r_0=\left(\pi/\sqrt{-2\lambda}\right)\xi$. Therefore the data plotted in Fig. \ref{fig:Lcomp}, i.e. $r_0/\left(\sqrt{2} \pi\right)$ describe a ``corrected healing length'' accounting for the modulation effect due to the variation of the (bound) ground state eigenvalue $\lambda$ across the BEC-BCS crossover. The factor $1/\sqrt{2}$ comes instead from a difference in the definition of $\xi$ with respect to \cite{THMuryshev}}. The values calculated in the framework of the EFT (black line) seem to be in good agreement with the results of the TDBdG equations (blue circles with error bars) across the whole range of available data. Moreover it seems that the present EFT approach captures well the fact that the characteristic length of the system changes from the healing length in the BEC regime (purple dot-dot-dashed line) to the correlation length in the BCS regime (green dashed line). In the far BEC limit the EFT results are in excellent agreement with both the data from \cite{THMunozBrand} (red dot-dashed line) and with the healing length obtained from the standard Gross-Pitaevskii treatment.
which can be therefore identified as the relevant length scale for the decay process. 
\\In Figure \ref{fig:kcZ} the effect of spin imbalance on the critical wavenumber for the instability is examined. It appears that the presence of unpaired particles stabilizes the soliton: the value of $k_c$ at a fixed interaction strength decreases when increasing the imbalance parameter $\zeta$, meaning that for a given width of the atomic cloud a soliton in an imbalanced setup can be stable while one in a balanced system is unstable. 
This can be qualitatively explained in terms of the observation that in an imbalanced configuration the soliton core is an energetically favorable place to accommodate the unpaired particles \cite{THLvAKTPrA}. Because of this the system may favor the soliton configuration over the vortex one since the former offers more space to store the excess component particles.

\section{Conclusions}\label{conclusions}
In this paper the snake instability mechanism for dark solitons in Fermi superfluids was studied in different conditions of interaction and imbalance. The theoretical treatment is based on an effective field theory \cite{THKTLDEpjB} already employed to discuss the properties of stable dark solitons in \textit{quasi-}1D configurations. 
The distortion of the depletion plane characteristic of the onset of the snake instability is treated by adding a transverse perturbation to the stable solution $\Psi_s$ of the equations of motion for the \textit{quasi-}1D configuration in the form \eqref{Phi}. The numerical solution of the coupled system of nonlinear differential equations describing the perturbation amplitude provides the spectra of the instability. The growth rate is examined in different interaction regimes and the BEC prediction \cite{THMuryshev} for the position of the maxima of $\text{Im}[\Omega(k)]$ is verified and extended to the BCS-side of the resonance. The minimal transverse size for the soliton decay is qualitatively estimated as $k_c^{-1}$, $k_c$ being the maximal wave number for which unstable modes exist. The results obtained show a good quantitative agreement with those of the coarse-grained BdG theory \cite{THMunozBrand} in the BEC-regime and the available numerical results of the TDBdG calculations \cite{THCetoliBrand} across the crossover. Moreover the EFT results seem to correctly characterize the change in the relevant length scale, from the condensate healing length in the BEC limit to the correlation length in the BCS regime.
\\ At a later stage in the paper the effects of spin-imbalance on the stability of the soliton are discussed. 
The maximum transverse size that the atomic cloud can have in order for the soliton to be stable is shown to increase in the presence of spin-imbalance. This could in principle offer a way to stabilize the soliton configuration in experiments without being forced to reduce the transverse size of the trap. The analysis carried out in the present paper is based on the perturbation of stable solitons solutions derived in \cite{THKTDPrA,THLvAKTPrA} which were obtained under the hypothesis of a uniform system. While most experiments concerning ultracold quantum gases employ harmonic traps to confine the atomic cloud, recently box-like optical traps that well approximate a uniform configuration were developed \cite{EXPGauntSchmidutz}: such setups can provide the opportunity of testing the predictions of this work in experiment. 

\acknowledgements We gratefully acknowledge useful discussions with J.P.A. Devreese,  and N. Verhelst. W. Van Alphen gratefully acknowledges support in the form of a Ph.D. fellowship of the Research Foundation-Flanders (FWO). This research was supported by the Flemish Research Foundation (FWO-Vl), project No. G.0115.12N, No. G.0119.12N, No. G.0122.12N, No. G.0429.15N, No. G0G6616N, by the Scientific Research Network of the Research
Foundation-Flanders, WO.033.09N, and by the Research Fund of the University of Antwerp.

\appendix
\section{coefficients of the EFT} \label{app}
\def\theequation{A.\arabic{equation}}
\setcounter{equation}{0}
In this section we give an overview of the coefficients appearing in the effective field action \eqref{Sbeta}.
It is convenient to write these coefficients in terms of the functions $f_j(\beta,\epsilon,\zeta)$, which are defined as the solutions of
\begin{align}
f_{j}(\beta,\epsilon,\zeta)=\frac{1}{\beta}\sum_{n}\frac{1}{\left[\left(\omega_n-i\zeta\right)^2+\epsilon^2\right]^j}
\end{align}
where $\omega_n$ are fermionic Matsubara frequencies of the form $\omega_n=(2n+1)\pi/\beta$.
The explicit expression for the first of these functions $f_1(\beta,\epsilon,\zeta)$ is given by
\begin{align}
f_1(\beta,\epsilon,\zeta)=\frac{1}{2\epsilon}\frac{\sinh(\beta\epsilon)}{\cosh(\beta\epsilon)+\cosh(\beta\zeta)}
\end{align}
From this, the other $f_j(\beta,\epsilon,\zeta)$ with $j=2,3,...$ can be calculated by using the simple recursion relation
\begin{align}
f_{j+1}(\beta,\epsilon,\zeta)=-\frac{1}{2j\,\epsilon}\frac{\partial f_j(\beta,\epsilon,\zeta)}{\partial \epsilon}
\end{align}
The complete expressions for the coefficients appearing in $S(\beta)$ \eqref{Sbeta} are hence given by
\begin{align}
C &  =\int\frac{d\mathbf{k}}{\left(  2\pi\right)  ^{3}%
}\frac{k^{2}}{3m}f_{2}\left(  \beta,E_{\mathbf{k}},\zeta\right)  ,\label{c}\\
D &  =\int\frac{d\mathbf{k}}{\left(  2\pi\right)  ^{3}}\frac
{\xi_{\mathbf{k}}}{w}\left[  f_{1}\left(  \beta,\xi_{\mathbf{k}},\zeta\right)
-f_{1}\left(  \beta,E_{\mathbf{k}},\zeta\right)  \right]  ,\label{d}\\
E  &  =2 \int\frac{d\mathbf{k}}{\left(  2\pi\right)  ^{3}}%
\frac{k^{2}}{3m}\xi_{\mathbf{k}}^{2}~f_{4}\left(  \beta,E_{\mathbf{k}}%
,\zeta\right)  ,\label{ee}
\end{align}
while the thermodynamic potential $\Omega_s$ reads
\begin{align}
\Omega_s(\Psi)=&-\int\frac{\mathrm{d}\bm{k}}{\left(2\pi\right)^3}\Bigg[\frac{1}{\beta}\log\left[2\cosh\left(\beta E_{\bm{k}}\right)+2\cosh\left(\beta \zeta\right)\right]+\nonumber\\&-\xi_{\bm{k}}-\frac{m\left|\Psi\right|^2}{k^2}\Bigg]-\frac{m\left|\Psi\right|^2}{4\pi a_s} \,.
\end{align}
The chemical potentials of the two pseudo-spin species $\mu_{\uparrow}$ and $\mu_{\downarrow}$ have been combined into the average chemical potential $\mu = (\mu_{\uparrow} + \mu_{\downarrow})/2$ and the  imbalance chemical potential $\zeta = (\mu_{\uparrow} - \mu_{\downarrow})/2$. This last parameter determines the difference between the number of particles in each spin-population. The quantity $\xi_{\bm{k}}=\frac{k^2}{2m}-\mu$ is the dispersion relation for a free fermion, $E_{\bm{k}}=\sqrt{\xi_{\bm{k}}+|\Psi|^2}$ is the single particle excitation energy and $a_s$ is the $s-$wave scattering length that determines the strength and sign of the contact interaction.
The coefficients $A$ and $\tilde{D}$ appearing in the equation of motion \eqref{eqofmot} are defined as
\begin{align}
A  &  =\frac{\partial\Omega_{s}\left(
\Psi\right)  }{\partial|\Psi|^2},\quad\tilde{D}
=\frac{\partial\left(  |\Psi|^2 D  \right)  }{\partial
|\Psi|^2}. \label{A&D}
\end{align}

\bibliography{Refs_experiment,Refs_theory}

\end{document}